\documentclass[twocolumn,superscriptaddress,aps,amsfonts,reprint,preprintnumbers,amsmath,amssymb,multicol]{revtex4}
\usepackage{wrapfig,subfig,graphicx,ulem}
\usepackage{xcolor}
\usepackage{dcolumn} 
\usepackage{bm} 

\newcommand\tb{\textbf}
\newcommand\bs{\boldsymbol}
\newcommand\p{\partial}
\newcommand\beq{\begin{equation}}
\newcommand\eeq{\end{equation}}

\begin{document}
\title{High-resolution measurements of the spatial and temporal evolution of megagauss magnetic fields created in intense short-pulse laser-plasma interactions} 
\author{Gourab Chatterjee}
\email{gourab@tifr.res.in}
\author{Prashant Kumar Singh} 
\author{Amitava Adak} 
\author{Amit D. Lad} 
\author{G. Ravindra Kumar}
\email{grk@tifr.res.in}
\affiliation{Tata Institute of Fundamental Research, 1 Homi Bhabha Road, Mumbai 400 005, India}

\begin{abstract}
A pump-probe polarimetric technique is demonstrated, which provides a complete, temporally and spatially-resolved mapping of the megagauss magnetic fields generated in intense short-pulse laser-plasma interactions. A normally-incident time-delayed probe pulse reflected from its critical surface undergoes a change in its ellipticity according to the magneto-optic Cotton-Mouton effect due to the azimuthal nature of the ambient self-generated megagauss magnetic fields. The temporal resolution of the magnetic field mapping is of the order of the pulsewidth, whereas a spatial resolution of a few microns is achieved by this optical technique. In addition, this technique does not suffer from refraction effects due to the steep plasma density gradients owing to the near-normal incidence of the probe pulse and consequently, higher harmonics of the probe can be employed to penetrate deeper into the plasma to even near-solid densities. The spatial and temporal evolution of the megagauss magnetic fields at the target front as well as at the target rear are presented. The micron-scale resolution of the magnetic field mapping provides valuable information on the filamentary instabilities at the target front, whereas probing the target rear mirrors the highly complex fast electron transport in intense laser-plasma interactions.        
\end{abstract}

\maketitle

\section{Introduction}
Irradiation of a solid with an intense ($\geq 10^{18}$ W/cm$^2$) short-pulse ($\leq 1$ ps) laser produces mega-ampere relativistic `fast' electron currents, which in turn lead to the generation of picosecond-duration pulses of megagauss magnetic fields \cite{KorobkinJETP1966, StamperPRL1971, StamperPRL1975, StamperPRL1978, StamperLPB1991, ThomsonPRL1975, RavenPRL1978, HainesCanJPhys1986, SudanPRL1993, HainesPRL1997, MasonTabak1998, BorghesiPRL1998, TatarakisNature2002, TatarakisPOP2002, WagnerPRE2004, GopalPOP2008, SandhuPRL2002, SandhuPRE2006, KahalyPOP2009, ChatterjeePRL2012, MondalPNAS2012, LiPRL2006, WillingalePRL2010, SarriPRL2012, SchumakerPRL2013}. In addition to serving as a crucial diagnostic in unraveling fast electron transport \cite{ChatterjeePRL2012, DaviesPRE2003}, with far-reaching implications in alternate ion-acceleration schemes \cite{SnavelyPRL2000} and the fast ignitor scheme of inertial confinement fusion \cite{TabakPOP1994}, these megagauss magnetic fields also provide an excellent platform for the simulation of astrophysical scenarios in the laboratory \cite{NilsonPRL2006}. 

Previous magnetic field measurements have employed an external transverse optical probe pulse, the Faraday rotation of which provides a measure of the megagauss magnetic fields \cite{StamperPRL1975, StamperPRL1978, RavenPRL1978, BorghesiPRL1998}. However, these measurements are confined only to the underdense plasma due to severe refraction from steep density gradients \cite{LehmbergPF1978}. An estimate of the magnetic fields at the critical density surface, where the largest magnetic fields are predicted to exist, is given by the $X$-wave cutoff of the self-generated harmonics \cite{TatarakisNature2002, TatarakisPOP2002, WagnerPRE2004, GopalPOP2008}. However, these measurements are limited by the pulsewidth of the main interaction pulse \cite{GopalPOP2008}. Magnetic fields have also been inferred from proton \cite{LiPRL2006, WillingalePRL2010, SarriPRL2012} as well as electron \cite{SchumakerPRL2013} deflectometry measurements, which provide time-resolved magnetic field information, but are spatially integrated over the target thickness. 

\begin{figure*}
\centering\includegraphics[width=0.85\textwidth]{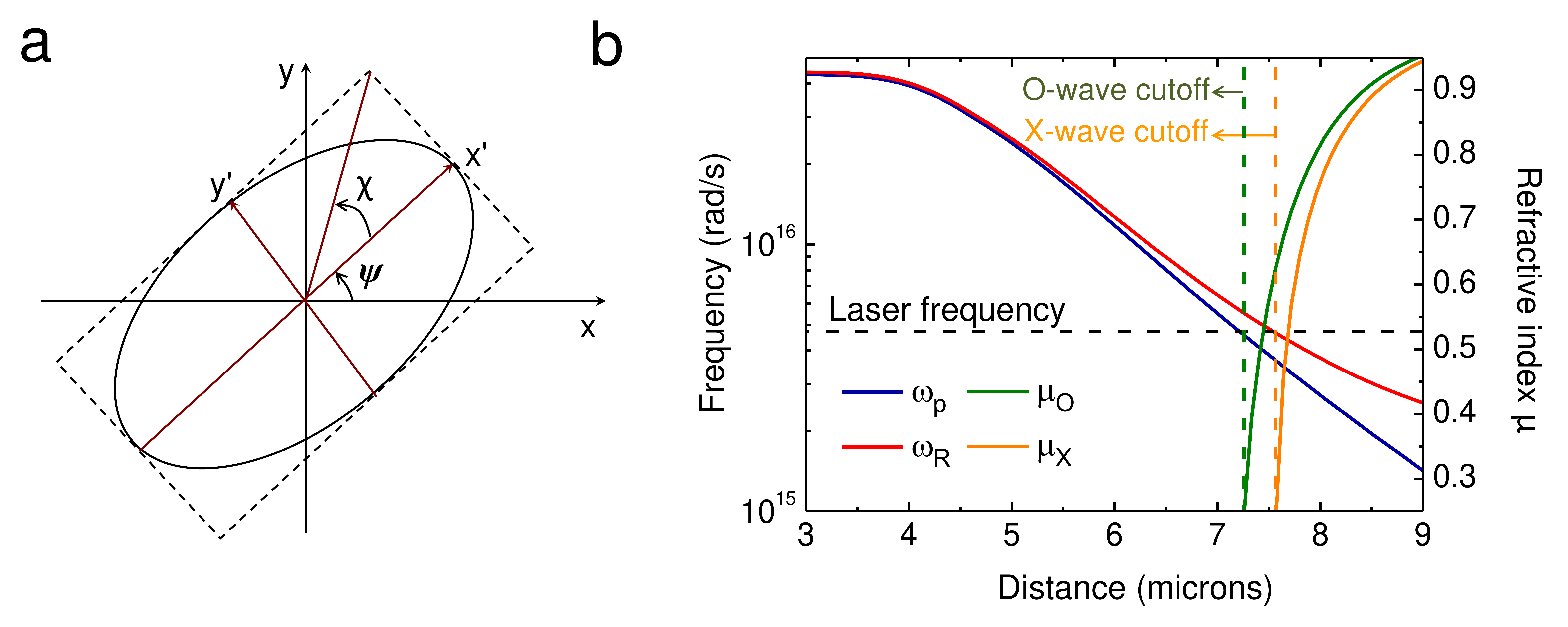}
\caption{\footnotesize{(a) The generalized polarization state of the reflected probe pulse with an induced ellipticity $\chi$ due to the azimuthal component of the magnetic field and a Faraday rotation $\psi$ due to the axial component of the magnetic field (which is usually negligibly small, given the predominantly azimuthal nature of the magnetic field). The incident probe pulse is linearly-polarized along the $x$ direction. (b) Schematic of the $O$-wave and $X$-wave cutoff. A typical electron density $n_e$ profile is mapped on to the corresponding plasma frequency $\omega_p$ profile (blue solid line) along with the right-hand $X$-wave cutoff $\omega_R$ (red solid line), assuming a uniform magnetic field of $B=100$ MG. The laser frequency for a probe wavelength of 400 nm is shown (black dashed line). The refractive indices $\mu_O$ (green solid line) and $\mu_X$ (orange solid line) are also plotted. The cutoff for the $O$- and $X$-waves (green and orange dashed lines) coincide with the points of intersection of the $\omega_p$ and $\omega_R$ curves with the laser frequency $\omega$, that is, with $\omega=\omega_p$ and $\omega=\omega_R$ respectively.}} 
\end{figure*}

This paper describes the systematics, implementation and physics applications of a pump-probe polarimetric technique for short-pulse intense-laser-produced plasmas, where the megagauss magnetic fields induce magneto-optic effects in the state of polarization of the probe pulse. Due to the azimuthal nature of the magnetic fields, the predominant change in the linearly-polarized normally-incident probe pulse is the introduction of ellipticity in its polarization state according to the Cotton-Mouton effect \cite{SegrePPCF1999,Hutchinsonbook}. The probe pulse can be time-delayed with respect to the main interaction pump pulse, with a resolution of the order of the probe pulsewidth. As a result, it is possible to monitor the temporal evolution of the megagauss magnetic fields, long after the interaction of the pump pulse with the solid. This is of pivotal importance in the magnetic field mapping at the target rear, a spatio-temporal study of which reveals the complex dynamics of fast electron transport that can last for several picoseconds after the laser-solid interaction \cite{ChatterjeePRL2012, DaviesPRE2003}. Moreover, a distinct advantage of this technique is that the probe pulse is affected by the magnetic fields only in the skin-depth of the plasma (since it reflects from its critical surface). Depending on the scale-length of the plasma electron density, this provides a highly localized mapping of the magnetic field along the longitudinal direction. For example, a magnetic field mapping at the target rear for different target thicknesses tracks the transport of fast electrons penetrating into the solid target after interaction at the target front surface \cite{GremilletPOP2002}. The transverse profile of the magnetic field also provides valuable information on fast electron transport and can be resolved with a precision of a few microns to mirror the dynamics of the various filamentary instabilities. Of crucial importance is the normal incidence of the probe pulse, which does not suffer from noticeable refraction effects due to steep plasma density gradients, unlike in oblique incidence \cite{LehmbergPF1978}. As a result, up-converting the probe pulse to its second or third harmonic allows it to probe deeper into the plasma. For instance, a third harmonic probe pulse is reflected from the highly overdense plasma at near-solid densities, almost an order of magnitude higher than the critical density of the main interaction pulse.

The paper is organized as follows: Section II describes the working principle of the Cotton-Mouton polarimetric technique employed for the measurement of the megagauss magnetic fields. Section III describes the pump-probe experimental setup as well as the multi-channel polarimetric measurements of the megagauss magnetic fields with a second-harmonic as well as a third-harmonic probe. The results are discussed in Section IV, which presents both frontside and rearside magnetic field measurements. Finally, the measurement technique as well as the experimental observations are summarized in Section V.

\section{Cotton-Mouton Polarimetry of Megagauss Magnetic Fields} 
The self-generated megagauss magnetic fields induce a birefringence in the plasma, resulting in a change in the state of polarization of the linearly-polarized incident probe pulse. For a normally incident probe pulse, the change in the state of polarization is manifested as a Faraday rotation $\psi$ due to the axial component of the magnetic field as well as the introduction of an ellipticity $\chi$ in accordance with the Cotton-Mouton effect due to the azimuthal component of the magnetic field \cite{SegrePPCF1999, Hutchinsonbook}, as shown in Fig. 1a.

In general, a complete description of the state of polarization of an electromagnetic wave is given by its Stokes' vector \cite{BornWolf}, defined as
\[
\tb{s}\equiv\begin{pmatrix} s_0 \\ s_1 \\ s_2 \\ s_3 \end{pmatrix}\equiv
I_0
\begin{pmatrix} 1 \\ \cos2\chi\cos2\psi \\ \cos2\chi\sin2\psi \\ \sin2\chi \end{pmatrix}.
\]

The state of polarization may also be represented by the Poincare sphere \cite{BornWolf, Goldstein}, which is a sphere of unit radius in the ($s_1$, $s_2$, $s_3$) space, where a given state of polarization is uniquely represented by a point on the Poincare sphere, with latitude $2\chi$ and longitude $2\psi$. The evolution of the Stokes' vector is then represented on the Poincare sphere by a rotation about an axis passing through the points representing the characteristic orthogonal polarization vectors and is thus described by the following evolution equation (along the $z$ direction) 

\begin{equation} 
\frac{d\tb{s}(z)}{dz}= \bs{\Omega}(z)\times \tb{s}(z),
\end{equation} 
where 
$$|\bs\Omega|\equiv\frac{\omega}{c}(\mu_O-\mu_X),$$ 
$\omega$ being the frequency of the incident laser pulse and $\mu_O$ and $\mu_X$  the refractive indices of the ordinary $O$- and the extraordinary $X$-waves respectively \cite{SegrePPCF1999, Hutchinsonbook}. 

The refractive index of the ordinary $O$-wave (neglecting collisions) is given by the usual relation
$$ \mu_O\equiv\sqrt{1-\frac{\omega_p^2}{\omega^2}}, $$
where $\omega_p$ is the plasma frequency, a function of the electron density $n_e$ and given by the relation $\omega_p^2\equiv 4\pi n_ee^2/m$, $e$ and $m$ being the charge and the mass of an electron respectively. On the other hand, the refractive index $\mu_X$ of the extraordinary $X$-wave depends on the ambient megagauss magnetic field $B$ via the cyclotron frequency $\omega_c=eB/mc$, according to the relation
$$ \mu_X\equiv\sqrt{1-\frac{\omega_p^2}{\omega^2}\left(\frac{\omega^2-\omega^2_p}{\omega^2-\omega^2_p-\omega_c^2}\right)}. $$
The different refractive indices lead to the accumulation of different phases by the two characteristic waves as the probe traverses across the plasma, which results in the induced ellipticity. 

The cut-off for the $O$-wave is at $\omega=\omega_p$, and hence it reflects from the usual critical density surface. However, the externally incident $X$-wave is reflected when $\omega=\omega_R$, where $\omega_R$ is the right-hand cutoff \cite{Chenbook}, given by the relation 
$$ \omega_R=\frac{1}{2}\left(\omega_c+\sqrt{\omega_c^2+4\omega_p^2}\right). $$
Thus, for a given incident laser frequency $\omega$, the $X$-wave reflects from a density lower than the critical surface density, from which the $O$-wave reflects \cite{Sandhuthesis}, as shown in Fig. 1b. The difference between the two depends on the ambient magnetic field $B$ and the scale-length of the interaction, which determines the spatial location of the two points of reflection.  

\section{Experimental Setup and Measurement of Megagauss Magnetic Fields}

\subsection{The pump-probe experimental setup}

\begin{figure}[t]
\centering\includegraphics[width=\columnwidth]{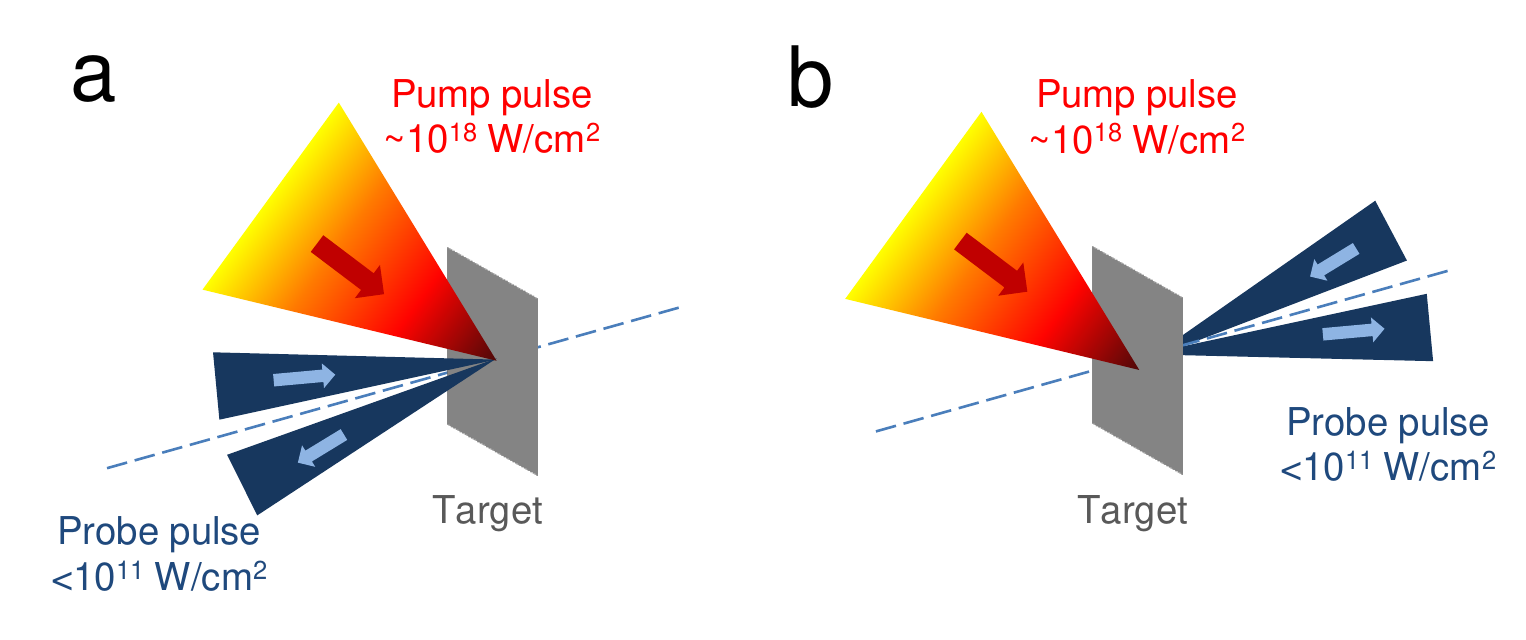}
\caption{\footnotesize{Schematic of the pump-probe experimental setup for probing the megagauss magnetic fields at the target (a) front and (b) rear.}} 
\end{figure}

The Tata Institute of Fundamental Research in Mumbai houses a 20 TW chirped-pulse-amplified Ti:Sa system, capable of producing 0.6 J, 30 fs pulses at a maximum repetition rate of 10 Hz. A schematic of the experimental setup is shown in Fig. 2. The $p$-polarized main interaction `pump' pulse has a pulsewidth of 30 fs and is centered at a wavelength of 800 nm. The pedestal 10 ps before the main interaction pulse has an intensity contrast of $10^{-5}$, whereas the prepulse leakage from the Pockels' cell has a nanosecond intensity contrast of $5\times 10^{-6}$ (see Fig. 5). An $f/3$ off-axis gold-coated parabolic mirror (typically with a focal length of $\sim$15 cm and an angle of incidence of $\sim25^0$) is optimized in order to determine the smallest focal spot, monitored by directly imaging the focal spot using a microscope objective. At the best focus, the focal spot has a diameter of 12 $\mu$m $\times$ 15 $\mu$m (FWHM). The angle of incidence of the main interaction pulse on the target is nearly 40$^0$. The target is mounted on a computer-controlled high-precision piezo-stage (Physik Instruments), which provides translational control in all the three directions with a precision of 0.018 $\mu$m as well as rotational control with a precision of $0.018^0$. The target is thus positioned at the focal plane of the main interaction laser pulse with a precision determined by the Rayleigh length of the focussing parabolic mirror. This is achieved by monitoring the hard x-ray emission due to the incidence of the main interaction pulse on the solid target, measured using a thallium-doped sodium-iodide scintillation detector, coupled with a photomultiplier and conventional electronic amplifiers.

A non-perturbing, well-attenuated `probe' pulse (typically at an intensity $<10^{11}$ W/cm$^2$) is extracted from the main interaction pump pulse using a thin beam-splitter. A temporal delay is introduced in the path of the probe pulse using a retro-reflecting mirror mounted on a computer-controlled delay stage (Akribis Systems), which can introduce a temporal delay of up to 500 ps with a precision of 3.3 fs (or equivalently 1 $\mu$m). A $\beta$-barium borate (BBO) crystal (Type I, $\theta=29.2^0$, $\phi=0^0$, 20 mm $\times$ 20 mm $\times$ 2 mm) is used to frequency-double the probe pulse. (See Section D for a third harmonic probe setup.) Not only can the second harmonic probe penetrate into the overdense plasma upto densities of 4$n_c$ ($n_c$ being the critical density of the main interaction pulse), but it also eliminates the noise due to pump irradiation, when the detectors are coupled with narrow-bandpass interference filters for the second harmonic. The probe is focused at near-normal incidence on the target (along the direction of the plasma density gradient), since a density gradient normal to the propagation direction introduces polarization-dependent phase shifts in the probe even in the absence of a magnetic field \cite{LehmbergPF1978}. The focusing geometry of the probe -- namely the focal spot size, optical magnification and resolution -- depends on the experimental requirement. Typically, the probe is focussed to a spot size, ranging from $\sim 50$ $\mu$m to a few hundred microns on the target, depending on whether it is sufficient to sample the plasma created by the pump pulse or whether it is required for the probe to capture the entire laser-excited region. The optical magnification is typically about $\sim 10$x, with a typical resolution of a few microns. The focusing geometry also depends significantly on the intervening optics between the BBO crystals and the point of focus (for example, lenses and glass windows of the vacuum chamber), which introduce a significant dispersion depending on the probe wavelength and consequently should either be minimized or accounted for. This is of particular importance in experiments using various probe harmonics simultaneously, or alternatively for measurements with the third harmonic ($\sim 266$ nm) probe, where the alignment is often done using a more `visible' second harmonic `pilot' beam. .

The pump and the probe focal spots, monitored by a microscope overlooking the focal region, are spatially overlapped on the target surface by imaging low-energy pump ablation with the probe. Alternatively, a fluorescence indicator (for the probe wavelength) also identifies the spatial location of the probe with respect to the pump. The spatial overlapping is achieved by adjusting a mirror mounted on a picomotor piezo-actuator in the path of incidence of the probe.

\begin{figure}[b]
\centering\includegraphics[width=\columnwidth]{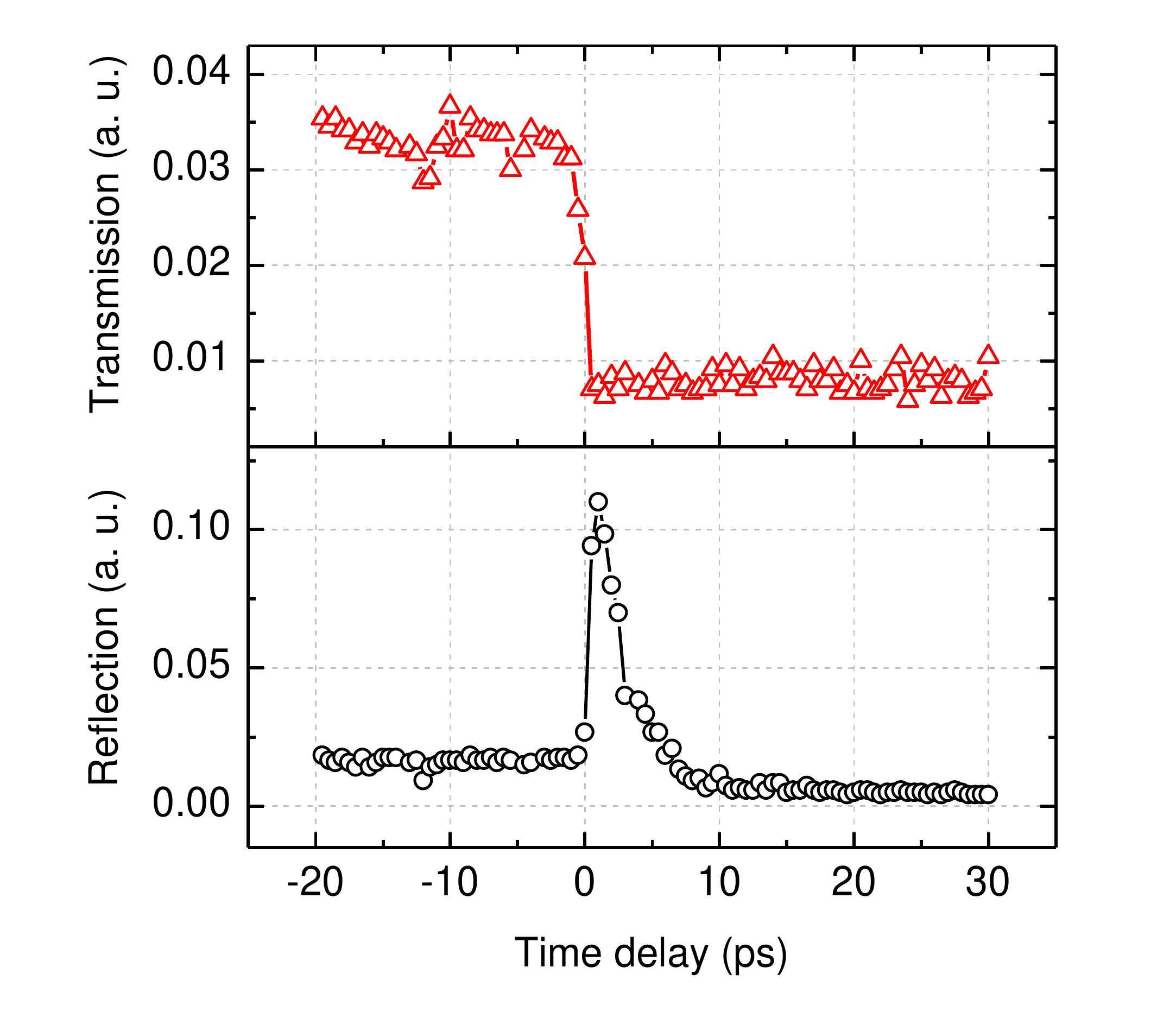}
\caption{\footnotesize{The reflected and transmitted probe signals (in arbitrary units), measured by photodiodes, indicating a sharp transition due to the reflection of the probe pulse from the plasma created by the main interaction pulse on a BK7 glass target and thereby determining the temporal overlap of the probe pulse with respect to the main interaction pump pulse. The irradiance is $3\times10^{16}$ W/cm$^2$ and the error is given by the shot-to-shot fluctuations at negative as well as very long probe delays.}} 
\end{figure}

\begin{figure}[t]
\centering\includegraphics[width=\columnwidth]{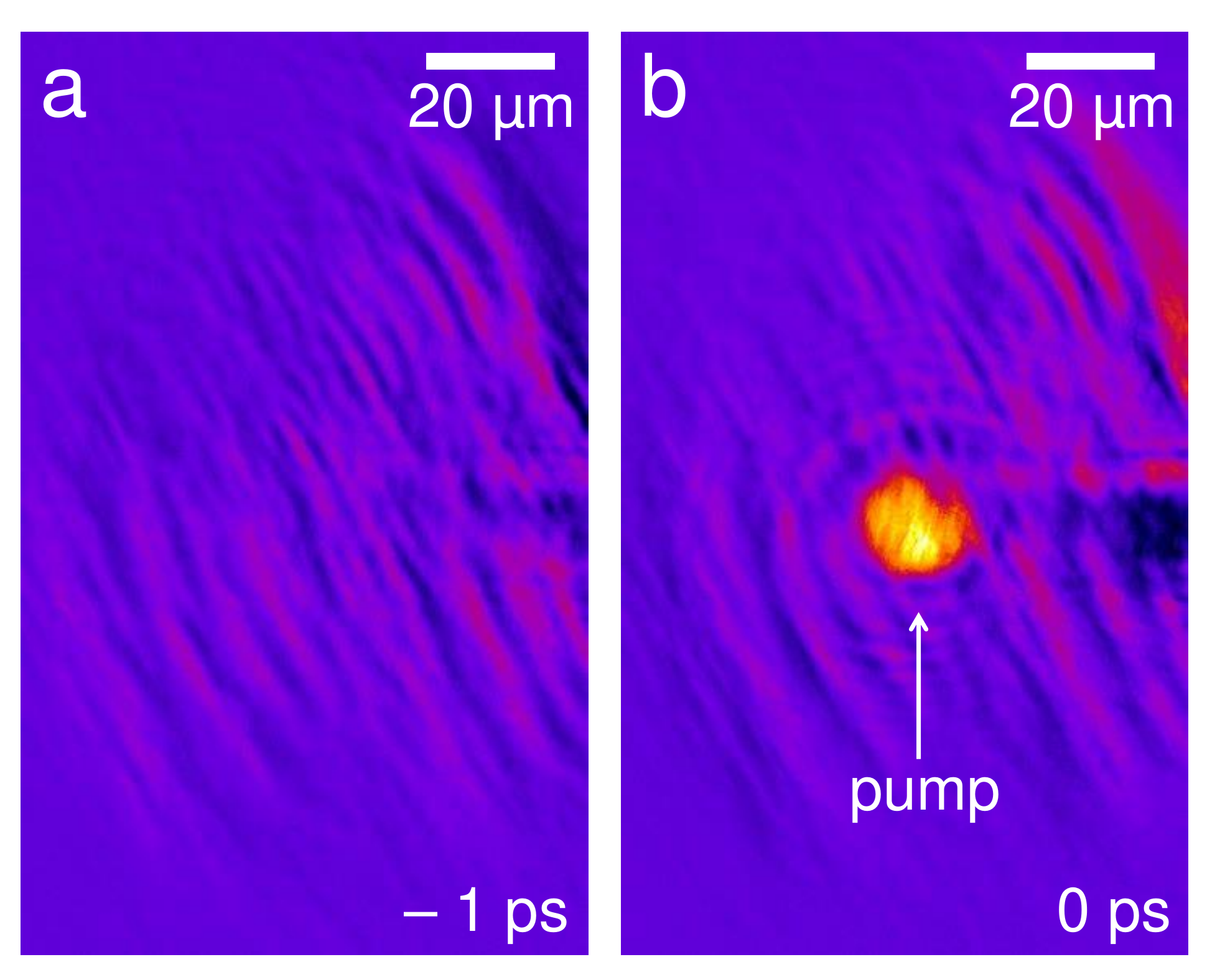}
\caption{\footnotesize{Spatially-resolved images of the reflected probe pulse from a BK7 glass target (a) 1 ps before the incidence of the main interaction pulse and (b) on the incidence of the main interaction pulse. A highly localized plasma formation can be clearly seen. The irradiance is $3\times10^{16}$ W/cm$^2$. The error in the temporal synchronization is $\pm 0.4$ ps, determined by the laser contrast.}} 
\end{figure}

The temporal synchronization of the pump and the probe pulses is achieved by varying the temporal delay of the probe with respect to the pump. This yields a sharp transition in the reflectivity as well as the transmissivity of the probe, signifying the reflection of the probe pulse from the plasma `mirror' created by the pump pulse, thereby determining the point of concurrent incidence of the two pulses. A typical example is shown in Fig. 3 at an irradiance of $3\times10^{16}$ W/cm$^2$. At a `negative' time-delay (probe preceding the pump pulse), the probe reflects from the solid target, which in this case is a polished BK7 glass slab. On the incidence of the interaction pump pulse (denoted as 0 ps in Fig. 3), the probe reflects from the plasma mirror formed, as a result of which there is a sudden rise in the probe reflectivity and a consequent drop in its transmissivity. The transition is distinctly sharp, initiated over time-scales of less than a picosecond. This is further demonstrated in Fig. 4b, where the spatially-resolved image of the probe beam reflected from a BK7 glass target shows a bright localized region of plasma formation (resulting in a high-reflectivity hot-spot) on the incidence of the main interaction pulse; even at a negative delay of 1 ps, there is no similar hot-spot (Fig. 4a). A quantitative estimate of the precision in the temporal synchronization is provided by the intensity contrast of the main interaction laser pulse, as shown in the third-order femtosecond cross-correlator (SEQUOIA) measurement trace in Fig. 5. The threshold intensity for plasma formation in BK7 glass is $3\times10^{13}$ W/cm$^2$ \cite{VonderlindeJOSAB1995} and consequently, a transition in the reflectivity/transmissivity might also correspond to preplasma formation due to a prepulse of intensity $\sim 10^{13}$ W/cm$^2$. Under moderate irradiance of $3\times10^{16}$ W/cm$^2$, the threshold intensity of plasma formation on BK7 glass corresponds to an intensity contrast of $10^{-3}$. As shown in Fig. 5b, such an intensity contrast is attained at a negative time-delay of 0.4 ps, which may be regarded as the error bar on all temporal measurements presented in this paper. 

For a normal-incidence geometry of the probe, where shadowgraphy/interferometry cannot be employed, the aforesaid technique of determining the temporal overlap between the pump and the probe pulses is highly convenient for high-repetition-rate femtosecond-pulsewidth laser systems. In addition, the transition in the self-reflectivity/transmissivity of the probe can also be employed to estimate the contrast of the laser system as it disappears with increasing intensity, depending on the contrast of the laser. 

\begin{figure}[t]
\centering\includegraphics[width=0.9\columnwidth]{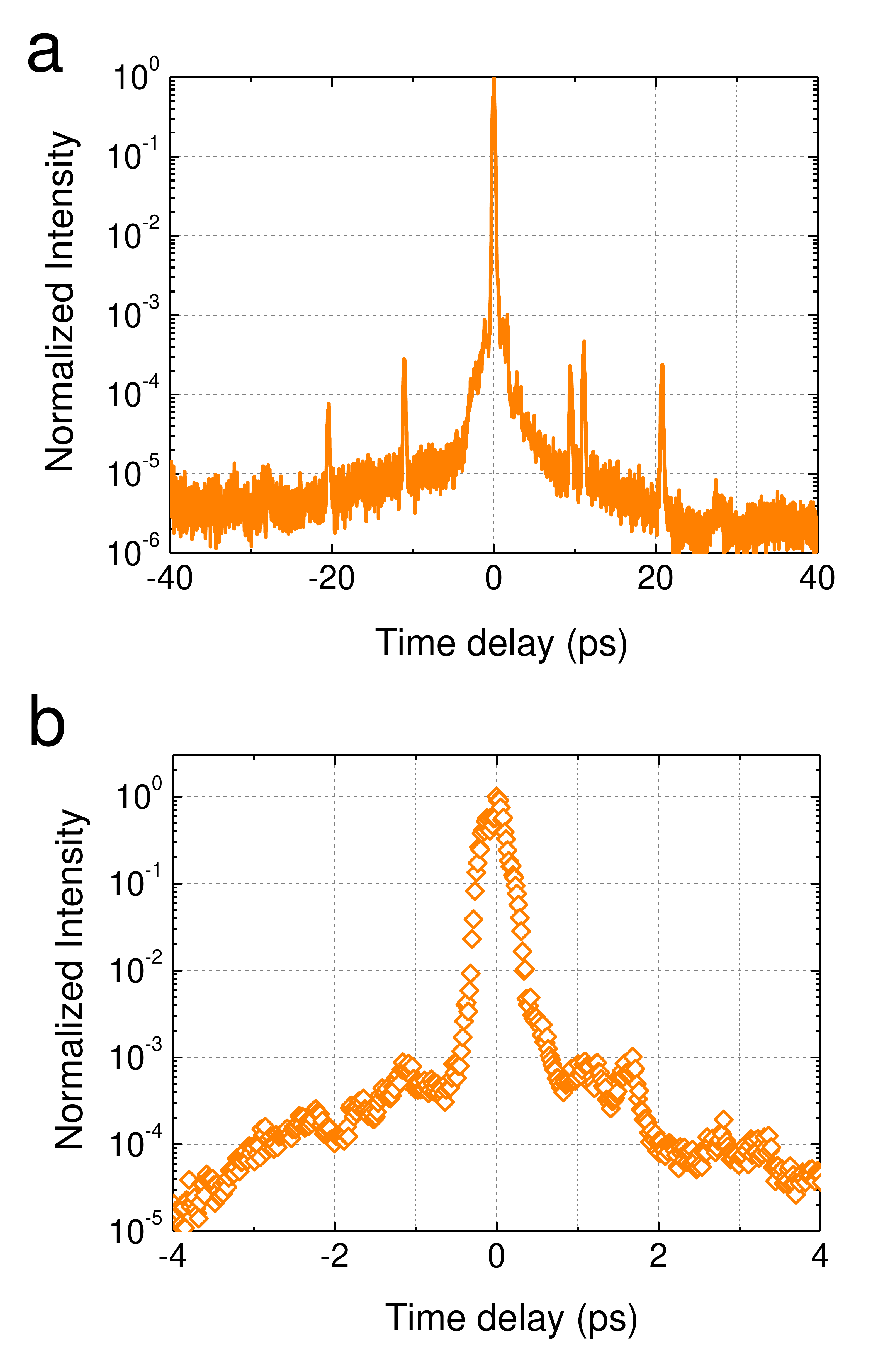}
\caption{\footnotesize{The intensity contrast of the main interaction pulse from (a) $- 40$ ps to $+ 40$ ps and (b) $-4$ ps to $+4$ ps (magnified for illustration), measured using a high-dynamic range third-order femtosecond cross-correlator (SEQUOIA). The additional peaks at an interval of 10 ps in (a) are artefacts of the measurement.}} 
\end{figure}

A closer inspection of Fig. 3 reveals two important features. Firstly, at longer time-scales, the probe reflectivity gradually decays due to absorption of the probe in the expanding plasma. Secondly, shot-to-shot fluctuations of the probe reflectivity/transmissivity at negative as well as at very long probe delays provide a good measure of the stability and shot-to-shot jitter of the laser energy itself.

\subsection{Multi-channel polarimetric setup to measure the induced ellipticity}

The ellipticity of the reflected probe is derived from its experimentally measured Stokes' vector according to the following procedure -- the magnitude of the reflected probe signal gives $I_0=s_0$; a polarizer with its polarization axis parallel to the incident probe polarization gives $I_1=I_0(1+s_1)/2$; a polarizer with its polarization axis at an angle of $45^0$ to the incident probe polarization gives $I_2=I_0(1+s_2)/2$; a quarter wave-plate, along with a polarizer at an angle of $45^0$ with respect to the quarter wave-plate, gives $I_3=I_0(1+s_3)/2$, where $s_1^2+s_2^2+s_3^2=s_0^2$ and hence, only three Stokes' parameters are independent \cite{SegrePPCF1999}. A mapping of all the Stokes' parameters provides the induced ellipticity $\chi$ and the Faraday rotation $\psi$ in the probe. However, the megagauss magnetic fields typically generated in intense laser-solid interactions are predominantly azimuthal in nature and therefore induce only an ellipticity in the normally-incident probe (the propagation vector $\tb{k}$ of the probe being perpendicular to the magnetic field $\tb B$). The Faraday rotation due to a negligibly small axial component of the magnetic field is usually below the threshold level of detection in the experiment. Under such circumstances, it is often sufficient only to measure the major and minor axes of the polarization ellipse (see Fig. 1a). In other words, two polarizers with the polarization axes parallel and perpendicular to the incident probe polarization yields the induced ellipticity, according to the experimental setup shown in Fig. 6. 

\begin{figure}[b]
\centering\includegraphics[width=\columnwidth]{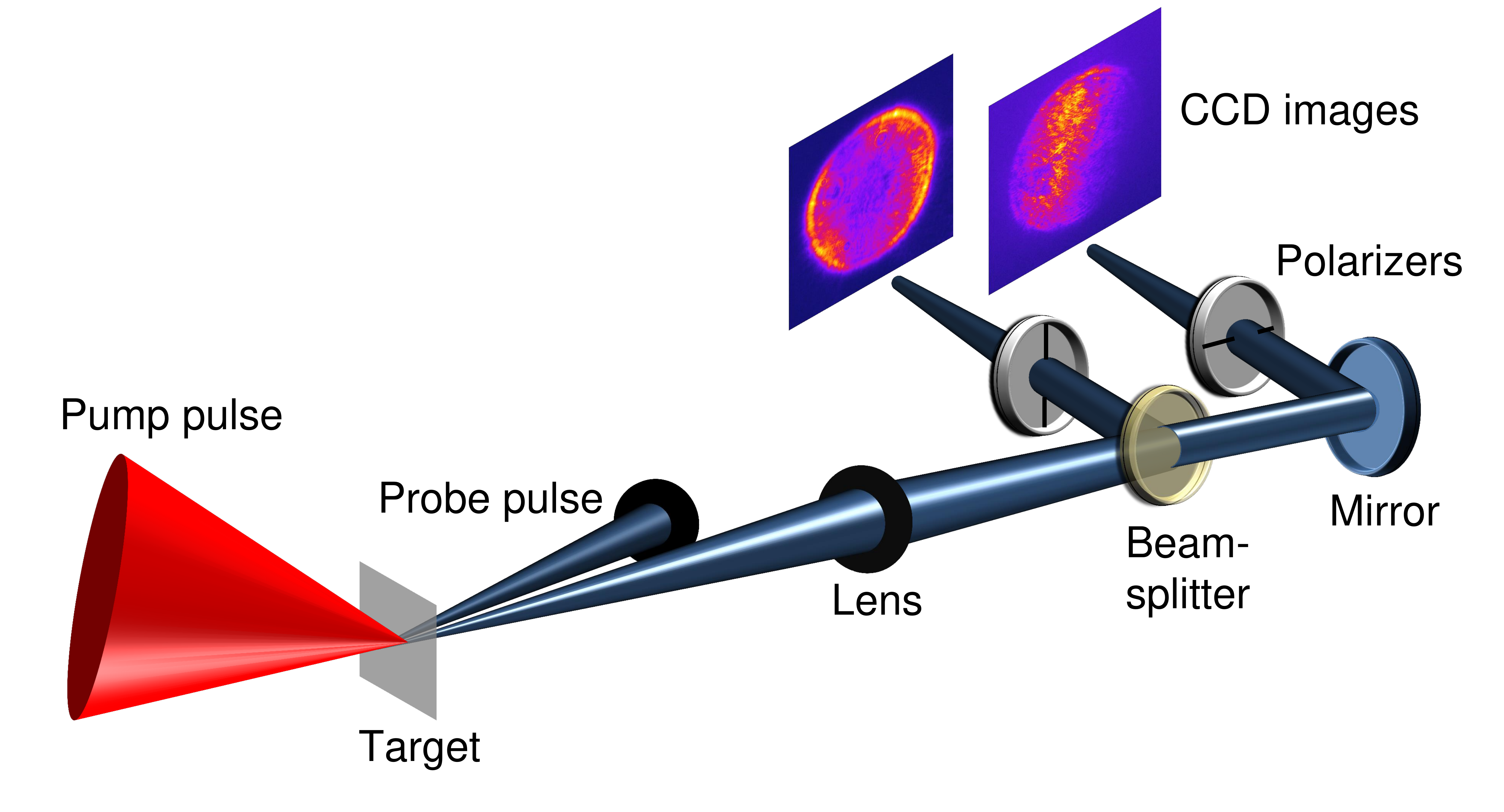}
\caption{\footnotesize{Schematic of the two-channel experimental setup for measuring the magnetic fields at the target rear. The reflected probe is split using a non-polarizing beam-splitter and directed into CCDs through polarizers. }} 
\end{figure}

The change in the state of polarization of the reflected probe pulse is analyzed using high-optical-quality Glan-air polarizers (Leysop), capable of producing a high extinction ratio of $10^{-6}$. Various optics in the path of the probe, particularly low-quality mirrors and beam-splitters, may induce depolarization and consequently, it is imperative to confirm whether the self-extinction of the probe using the polarizer is at least of the order of $10^{-4}$ before commencing on the experiment. The reflected probe signal may be measured using callibrated photodiodes for a time-resolved but spatially-integrated measurement. However, for a complete spatio-temporal mapping, we use high-dynamic-range mega-pixel charge-coupled devices (CCDs) (PI-Acton CoolSnap EZ, $1392\times1040$ pixels, pixel size 6.45 $\mu$m $\times$ 6.45 $\mu$m). The background to the signal in the CCD provides a measure of the ambient noise as well as the dark current of the CCD. In addition, the similar ellipticity profiles at negative time-delays with a near-zero magnitude provide a point of reference, and indicate that any induced ellipticity at positive time-delays can solely be attributed to the magnetic field generated on the incidence of the main interaction pulse. Besides, during the evaluation of the Stokes' parameters, the experimentally measured intensities are effectively normalized with respect to the reflected probe beam profile and consequently, the ellipticity estimated is independent of the local spatial fluctuations and intensity modulations in the probe beam profile. This is of particular importance in state-of-the-art high-intensity lasers, where shot-to-shot fluctuations can be a significant source of error. 

\subsection{Estimation of magnetic fields from experimentally measured ellipticity}

The magnetic field is estimated from the experimentally measured ellipticity by numerically solving the evolution equation for the magnetic field (see Eq. 1 above in Section II). The probe incident on the vacuum-plasma interface is linearly-polarized. As the probe penetrates into the plasma and reflects off its critical surface, the length traversed by the probe in the plasma is simulated as a plasma box divided into smaller cells, where the output Stokes' vector of one cell is fed as the input Stokes' vector of the adjacent cell. The evolution of the Stokes' vector is therefore monitored over the entire trajectory of the probe pulse, and the total ellipticity induced in the initially linearly-polarized probe is calculated. Corroborating the experimentally measured ellipticity with the calculated ellipticity yields the magnetic field. 

The ellipticity profile, derived from the polarization changes in the reflected probe imaged on to the CCDs and carrying the magnetic field information, is then used to decode the magnetic field profile on a pixel-by-pixel basis. For megapixel cameras, this technique can be computationally exhaustive. A convenient algorithm is to obtain the corresponding magnetic field for a given value of ellipticity up to the required degree of precision, typically 1 MG (in other words, ellipticity values yielding magnetic field values differing by less than 1 MG are not distinguished). The computation is then repeated for a list of ellipticity values and this conversion table (interpolated, if required) is used to obtain the magnetic field profile from the experimentally measured ellipticity profile pixel-by-pixel. Since typical values of ellipticity are in the range $10^{-3}-10^{-1}$ for megagauss magnetic fields, the computation is done for $\sim 10^3$ values of ellipticity instead of $\sim 10^6$ (corresponding to a megapixel ellipticity profile). Calculating the magnetic field at a longer time-delay requires significantly more processing time and the computation may be done for a lesser number of data points, coupled with suitable interpolation using appropriate frequency sampling \cite{uphill}.

In order to infer the magnetic field from the induced ellipticity, it is necessary to estimate the scale-length of the interaction, which determines the spatial location of the critical and cutoff surfaces for the $O$- and $X$-waves respectively. This can be directly measured by the transverse probing of the critical surface expansion using shadowgraphy or interferometry \cite{TatarakisPRL1998}. Alternatively, the plasma density profile can be modelled by assuming that the plasma expands into the vacuum at approximately the ion acoustic speed $c_s$ in a self-similar fashion \cite{MoraPRL2003}. The expansion velocity can be experimentally measured by monitoring the Doppler shift induced in the frequency of the probe pulse due to the motion of the critical surface from which the probe reflects \cite{MondalPRL2010}. The validity of the above measurements has been amply demonstrated by extensive hydrodynamic modelling and simulations \cite{LancasterPOP2009, MondalPRL2010, GremilletPOP2002, GMalkaPRE2008}.  

\subsection{Probing near-solid densities with a third harmonic probe}

A third-harmonic probe can access densities as high as 9$n_c$ ($\sim 10^{22}$ /cm$^3$) on normal incidence, where $n_c$ is the critical density for the main interaction pulse. In other words, it is possible to probe the highly overdense plasma at near-solid densities by upconverting the probe to its third harmonic.

For a Ti:Sa laser system, the third harmonic probe is centered at a wavelength of 266 nm and can be generated using a pair of BBO crystals. The first BBO crystal (Type I, $\theta=29.2^0$, $\phi=0^0$, 20 mm $\times$ 20 mm $\times$ 0.2 mm) generates the second harmonic (400 nm) with an efficiency of $\sim 25\%$, in addition to the residual fundamental frequency (800 nm). A second BBO crystal (Type I, $\theta=44.3^0$, $\phi=0^0$, 20 mm $\times$ 20 mm $\times$ 0.2 mm) can then be employed to yield the third harmonic (266 nm) from the second harmonic and the fundamental frequency via sum frequency generation (SFG) with an efficiency of (2-3)\%. 

A higher percentage yield of the third harmonic may be obtained by the following technique \cite{WangAPL2010}. A thick ($\sim 2$ mm) second-harmonic-generating BBO may be employed for a higher conversion efficiency of the second harmonic. However, it introduces a significant group velocity mismatch (GVM) between the generated second harmonic and the residual fundamental frequency, as a result of which the effective third harmonic yield (on using another BBO crystal) is significantly reduced. In this case, the GVM needs to be compensated by an additional BBO crystal (Type I, $\theta=29.2^0$, $\phi=0^0$, 20 mm $\times$ 20 mm $\times$ 0.2 mm), the crystallographic axis of which needs to be rotated by an angle of 45$^0$ with respect to that of the second-harmonic-generating BBO. Following the efficient generation of the second harmonic with compensated GVM, the third harmonic can be effectively generated using a third BBO crystal, as described above. 

All reflective optics in the probe imaging setup need to be converted to ultraviolet(UV)-enhanced aluminum-coated mirrors and all transmissive optics (for example, lenses and beam-splitters) need to be constituted of a UV-transparent medium such as fused silica or quartz (BK7 glass has a transmission range of 380-2100 nm). The signal is obtained in a CCD sensitive to UV radiation (Andor iKon-M 934 with UV-sensitive BU2 coating, 1024 $\times$ 1024 pixels, pixel size 13 $\mu$m $\times$ 13 $\mu$m) coupled with appropriate narrow-bandpass interference filters.

\section{Results and Discussions}

\begin{figure}[b]
\centering\includegraphics[width=\columnwidth]{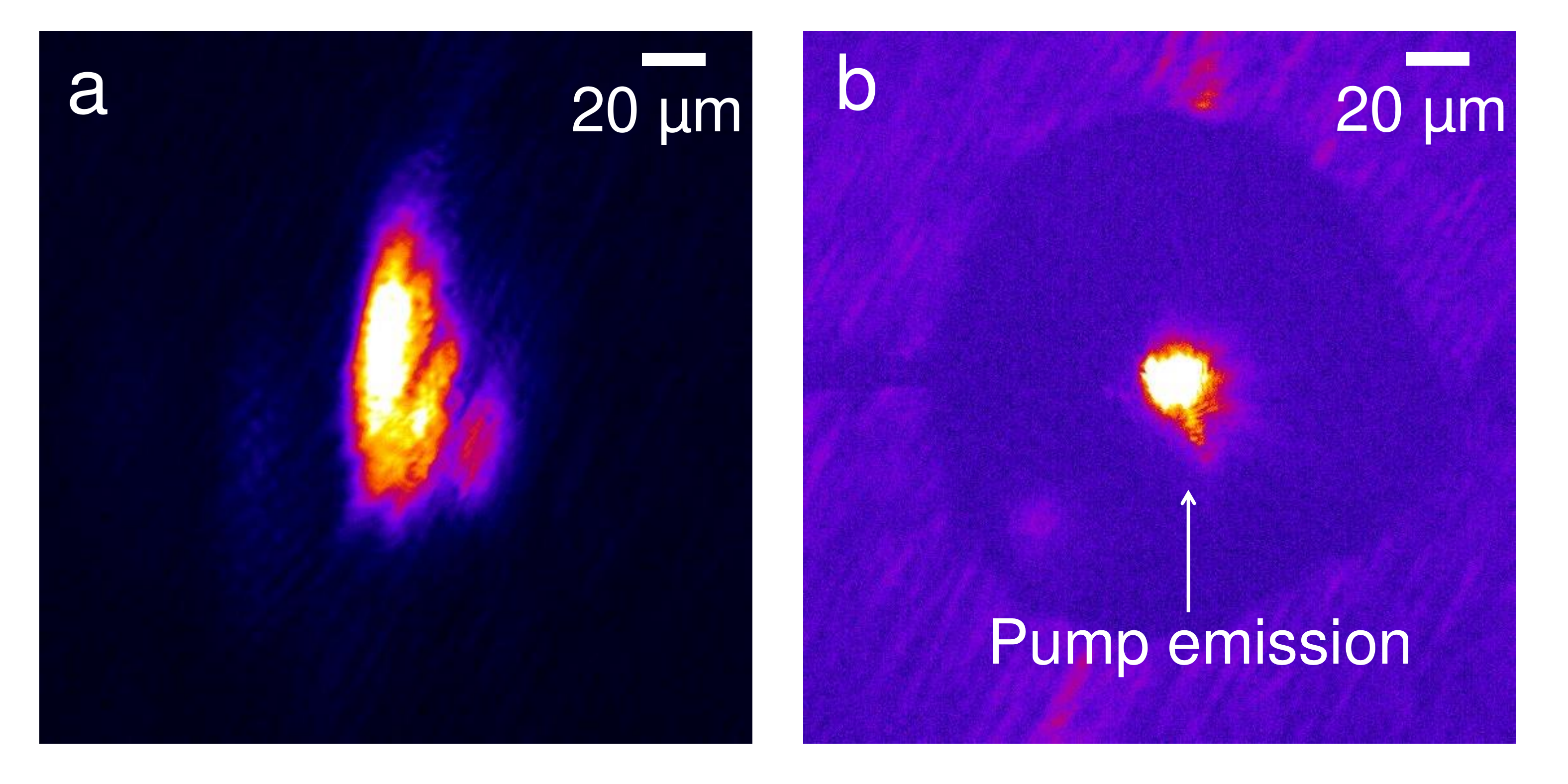}
\caption{\footnotesize{Second-harmonic signal on the CCD coupled to the crossed polarizer for the incidence of (a) probe only and (b) both pump and probe at a temporal delay of 50 ps in an aluminum target at an irradiance of $3\times10^{18}$ W/cm$^2$ despite a narrow-bandpass interference filter at the probe wavelength. The pump emission completely masks the probe, which is also significantly absorbed by the plasma, particularly at delays of $\sim 50$ ps. In fact, the pump emission saturates the CCD under filter conditions of (a) and hence additional neutral-density filters are required to obtain (b). (The probe was not made aberration-free for this test shot.)}} 
\end{figure}

\begin{figure}[t]
\centering\includegraphics[width=\columnwidth]{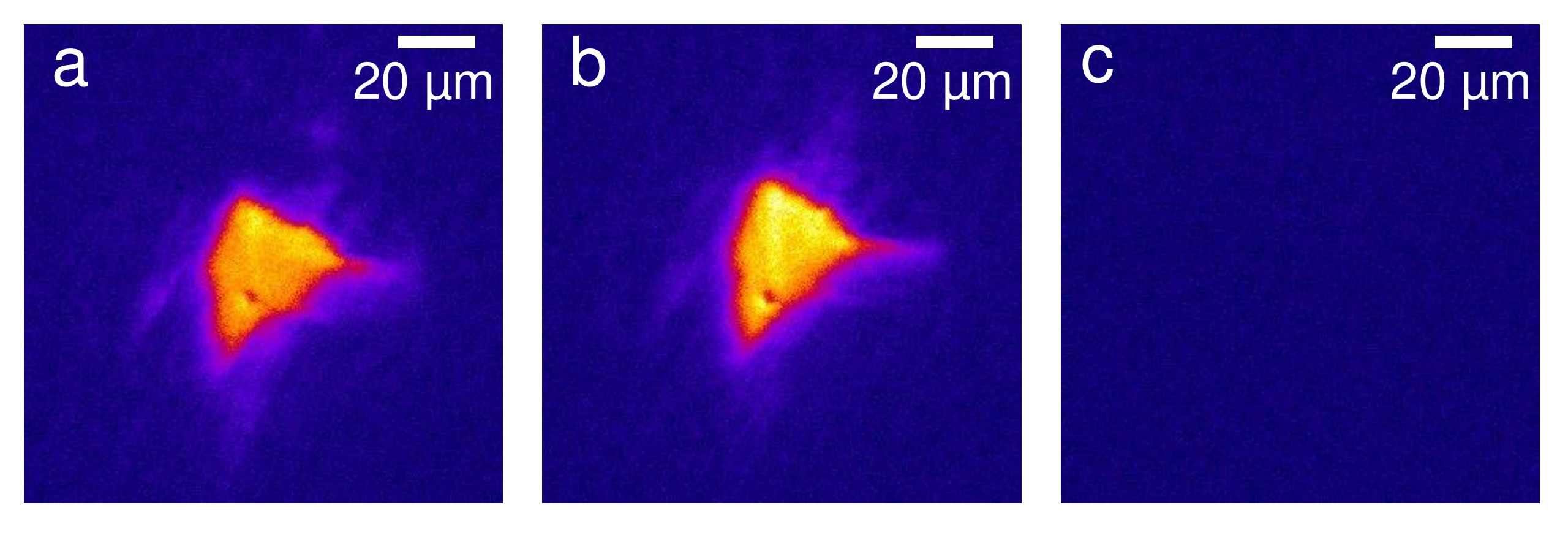}
\caption{\footnotesize{Second-harmonic signal on the ICCD coupled to the crossed polarizer for the incidence of (a) both pump and probe, (b) probe only and (c) pump only for an aluminum target, thereby indicating that there is no pump-noise even at an intensity of $3\times10^{18}$ W/cm$^2$ on using appropriate interference filters and temporally gating the signal with a nanosecond exposure on the ICCD. The filter conditions are the same for (a)-(c). (The probe was not made aberration-free for this test shot.)}} 
\end{figure}

For measurements at the target front, the reflected probe signal is often masked by a host of other sources of extraneous noise from pump pulse irradiation at the highest intensities. Of crucial importance is the ability to distinguish the probe in the CCD from the pump noise, particularly in the signal from the crossed polarizer, where the signal level is considerably low and consequently the CCD is coupled with very few or no neutral-density filters. This results in a high degree of unattenuated ambient radiation, which is often channeled through the imaging optics and is therefore focused at the image of the point of interaction on the CCD. An example is shown in Fig. 7, where despite narrow-bandpass interference filters, the weak probe signal from the crossed polarizer at a temporal delay of 50 ps is masked completely by the pump emission for an incident irradiance of $3\times10^{18}$ W/cm$^2$. At longer delays in particular, the probe signal is considerably weaker due to absorption in the plasma, which is even more pronounced for metallic targets with low ionization thresholds and consequently long scale-lengths of interaction. The pump noise may be predominantly attributed to (i) the scatter of the self-generated laser-harmonics collected by the imaging optics even along the direction of the target normal and (ii) broad-band plasma emission. In addition, metallic targets may often have strong emissions at these optical frequencies. For example, copper (Cu I) has strong line-emissions at 402 nm as well as 406 nm \cite{NIST}, which cannot be eliminated despite using narrow-bandpass (typically 10 nm bandwidth) interference filters at the probe wavelength of 400 nm \cite{Doppler}. Besides, plasma emission lasts until long after the interaction (typical CCDs have an exposure window of microseconds) and even if captured in the small spectral window allowed by the interference filter, still contributes significantly to the pump noise. Replacing the CCDs with intensified CCDs (ICCDs) having a nanosecond exposure window \cite{Pockels} eliminates both the plasma emission, which peaks in the visible window typically a few nanoseconds after the interaction, as well as most spectral line-emissions in the visible regime. Self-generated harmonics of the main interaction pulse from the plasma are typically of the duration of the laser pulsewidth and cannot be eliminated even by the usage of ICCDs. However, laser harmonics are polarized and are consequently strongly attenuated by the crossed polarizer, which is most susceptible to pump noise due to the low signal level. Another alternative is to use an anharmonic probe, typically obtained by Raman-shifting a harmonic probe, along with appropriate interference filters. Figure 8 shows a scenario where practically all the noise due to pump irradiation has been eliminated in the probe signal even at the highest intensities by a careful alignment of the probe path and the usage of an ICCD coupled with appropriate interference filters. 

\begin{figure}[t]
\centering\includegraphics[width=\columnwidth]{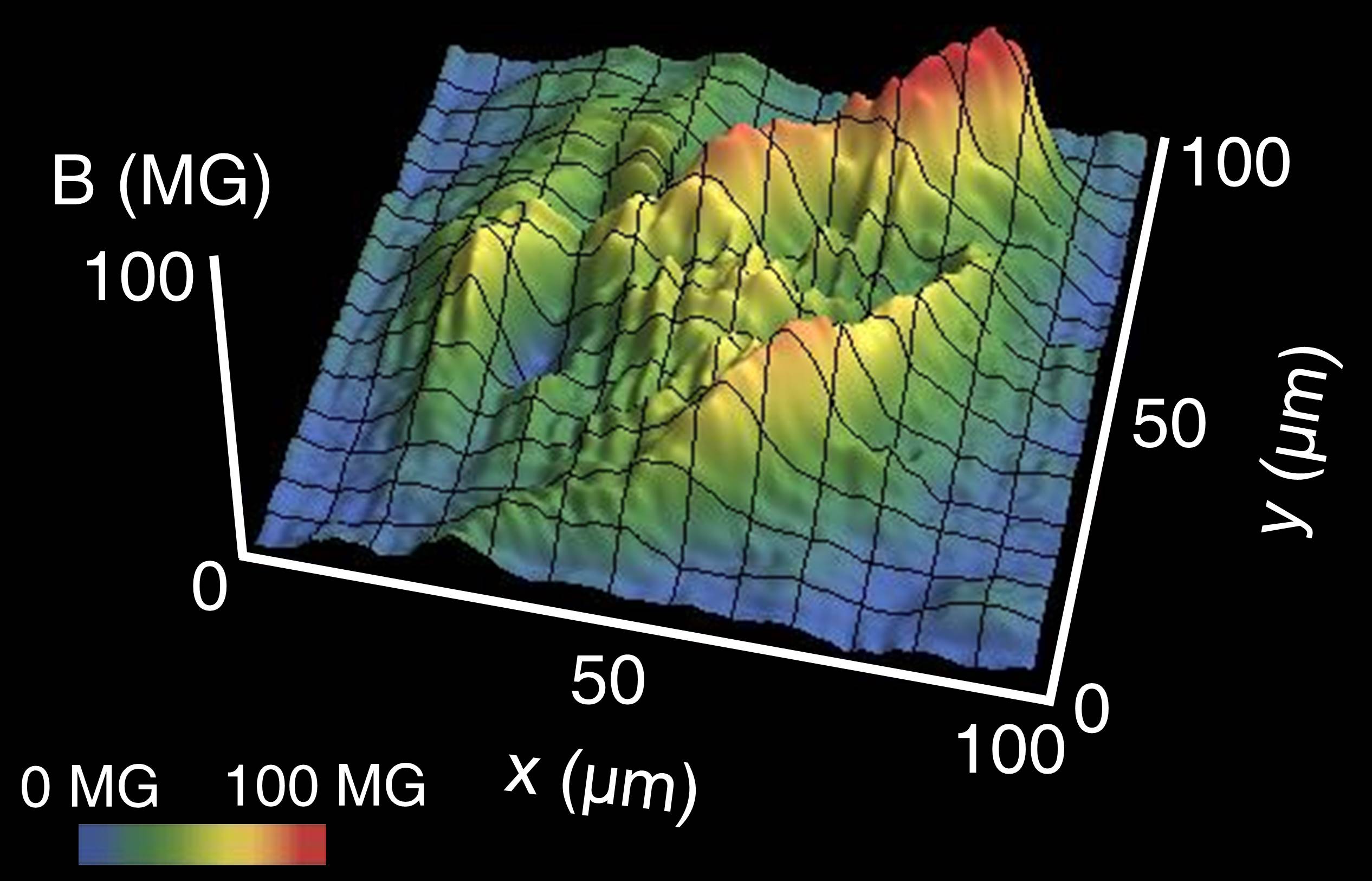}
\caption{\footnotesize{Spatial profile of the magnetic field at the front of an aluminum target at a temporal delay of 1 ps, obtained using ICCDs. The profile shows distinct filamentary structures and reaches local peak values of about 100 MG.  The intensity is $4\times 10^{18}$ W/cm$^2$.}} 
\end{figure}

Measurements at the target rear are reasonably devoid of pump noise (compared to measurements at the target front). Using an ICCD, one can eliminate plasma emission and for a normally incident probe at the target rear, the chief source of noise collected by the imaging optics along the target normal direction is optical transition radiation \cite{SantosPRL2002}, which is usually weaker than the probe for a thick (few tens of microns) target. 

Figure 9 shows a typical magnetic field profile at the target front at a temporal delay of 1 ps for an incident intensity of $4\times10^{18}$ W/cm$^2$ using a second harmonic probe. The magnetic field reaches typical peak values of $\sim 100$ MG and exhibits a high degree of filamentation. The origin of the filamentary structures in the magnetic field may be attributed to a plethora of mechanisms, most notably Weibel-type instabilities \cite{WeibelPRL1959}, which arise as the fast electron currents are spatially interspersed with the cold return currents they induce in the background plasma \cite{GremilletPOP2002}. Although the magnetic field polarigrams exhibit complete randomness in the distribution of the filamentary structures, in a recent experiment \cite{MondalPNAS2012} we showed that the Fourier spectra of the magnetic energy display a power-law scaling, which is a distinctive signature of magnetic turbulence \cite{FalgaronePassotbook}. In light of the ubiquity of turbulence (ranging from astrophysical scenarios to macroscopic terrestrial fluid motions) \cite{FalgaronePassotbook}, our results seem even more fascinating as they portray dynamic turbulent mechanisms in highly non-equilibrium regimes.

\begin{figure*}
\includegraphics[width=\textwidth]{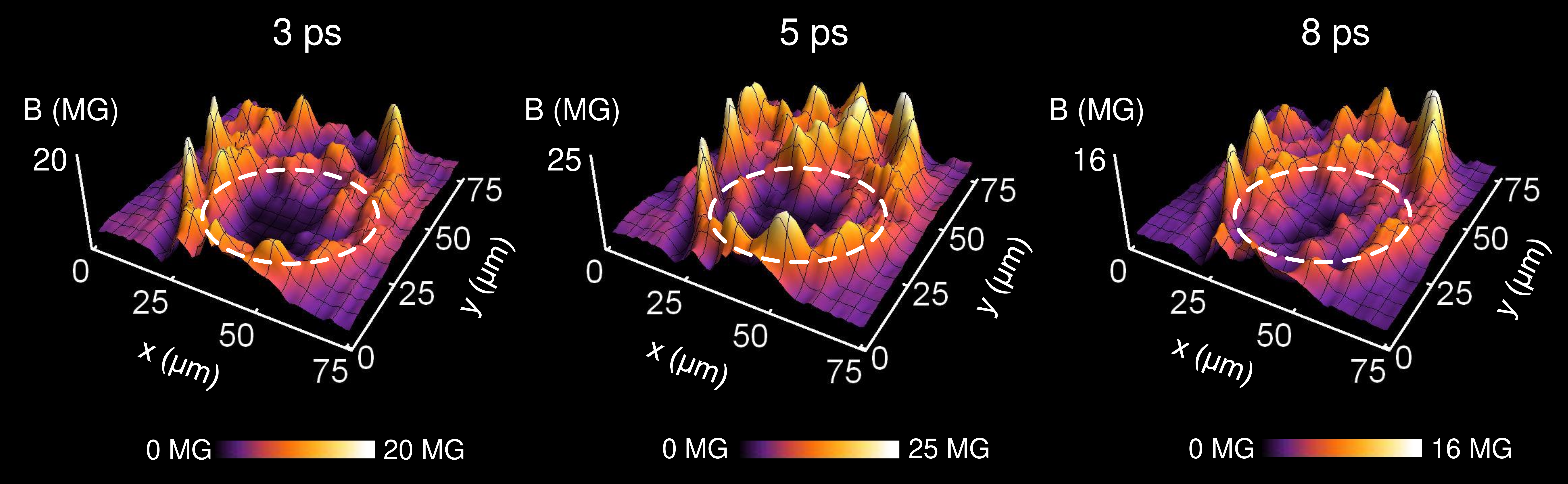}
\caption{\footnotesize{Spatial profile of the magnetic field at the rear of a 100 $\mu$m thick fused silica target at temporal delays of 3 ps (left), 5 ps (middle) and 8 ps (right) for an incident intensity of $4\times10^{18}$ W/cm$^2$. The annular nature of the magnetic field for the first few picoseconds indicates a beam-like fast electron transport. However, at later time-scales ($\sim$ 8-10 ps) the distinct hollow disappears (highlighted by the dashed white circle), possibly due to beam fragmentation.}} 
\end{figure*}

Most notably, in Fig. 9 there is no hollow in the magnetic field profile, unlike the magnetic field at the target rear, shown in Fig. 10. It is well-known that self-generated magnetic fields produce a collimating effect on the fast electron currents \cite{DaviesPRE2003}. As a result, the magnetic field at the rear of a 100 $\mu$m thick fused silica target is annular in shape (Fig. 10), thereby indicating a beam-like electron transport. In a recent experiment \cite{ChatterjeePRL2012} we demonstrated how the annular magnetic field profile measured at the rear of a sandwich target comprising 1.1 mm thick aligned carbon-nanotubes and 0.1 mm thick fused silica was qualitatively similar to the magnetic field profile at the rear of the 0.1 mm thick fused silica target alone, although the magnetic field at the rear of the sandwich target was found to be nearly an order of magnitude higher. This indicated a collimated and a highly efficient fast electron transport across the aligned carbon nanotube arrays over millimeter distances. These results were also corroborated by hot electron spectrum measurements at the rear of the aligned carbon nanotubes. 

Figure 10 shows the temporal evolution of the annular magnetic field profile, the hollow being most prominent for the first few picoseconds, and gradually disappearing at longer time-scales, possibly due to beam-breakup. Similar results were also obtained in our recent experiments \cite{ChatterjeeArxiv} at the Vulcan Petawatt laser facility for an irradiance of $4\times10^{20}$ W/cm$^2$. The magnetic field profile at the rear of a 50 $\mu$m thick plastic target exhibited an annular profile at a temporal delay of 5 ps, followed by a fragmentation of the collimated electron beam at later time-scales ($\sim 10$ ps).

\begin{figure}[b]
\centering\includegraphics[width=0.9\columnwidth]{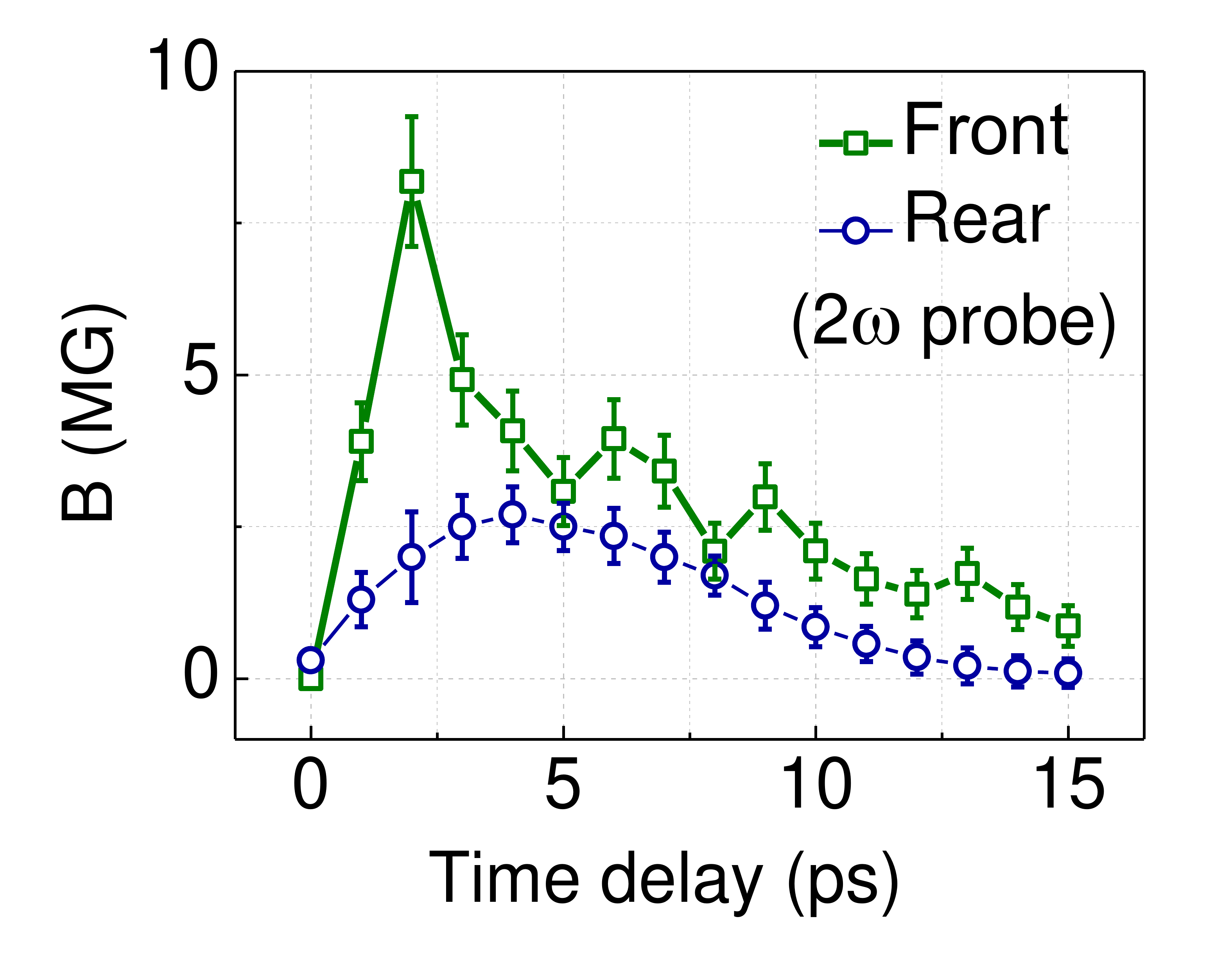}
\caption{\footnotesize{Temporal evolution of the spatially-averaged (across the transverse profile) magnetic field at the front as well as the rear of a 100 $\mu$m thick fused silica target at an incident intensity of $3\times10^{18}$ W/cm$^2$ using a second harmonic probe.}} 
\end{figure}

Figure 11 shows the temporal evolution of the megagauss magnetic field (spatially-averaged across the transverse profile) at the front as well as at the rear of a 100 $\mu$m thick polished fused silica target at an intensity of $3\times10^{18}$ W/cm$^2$. The spatially-averaged magnetic field at the target front reaches peak values of $\sim 10$ MG, although localized spatial fluctuations can be as high as $\sim 100$ MG, as indicated in Fig. 9. The growth and the decay of the magnetic field $\tb B$ may be expressed by the following analytical approximation (neglecting thermo-electric terms etc.), which essentially follows from Ohm's law and Maxwell's equations:
$$ \frac{\p \tb B}{\p t}=\nabla\times\left(\eta \tb{j}_f\right)-\nabla\times\left(\frac{\eta}{\mu_0}\nabla\times\tb B\right), $$
where $\eta$ is the resistivity of the background plasma and $\tb j_f$ is the forward current due to the fast electrons. Thus, the above equation can be distinguished into a source term and a diffusion term, both of which crucially depend on the resistivity $\eta$ and its gradient, that govern the rise and fall time-scales of the magnetic field. A detailed enumeration of the various processes leading to the time-evolution of the magnetic field may be found in the seminal work of Haines \cite{HainesCanJPhys1986}. 

\begin{figure}[b]
\centering\includegraphics[width=0.9\columnwidth]{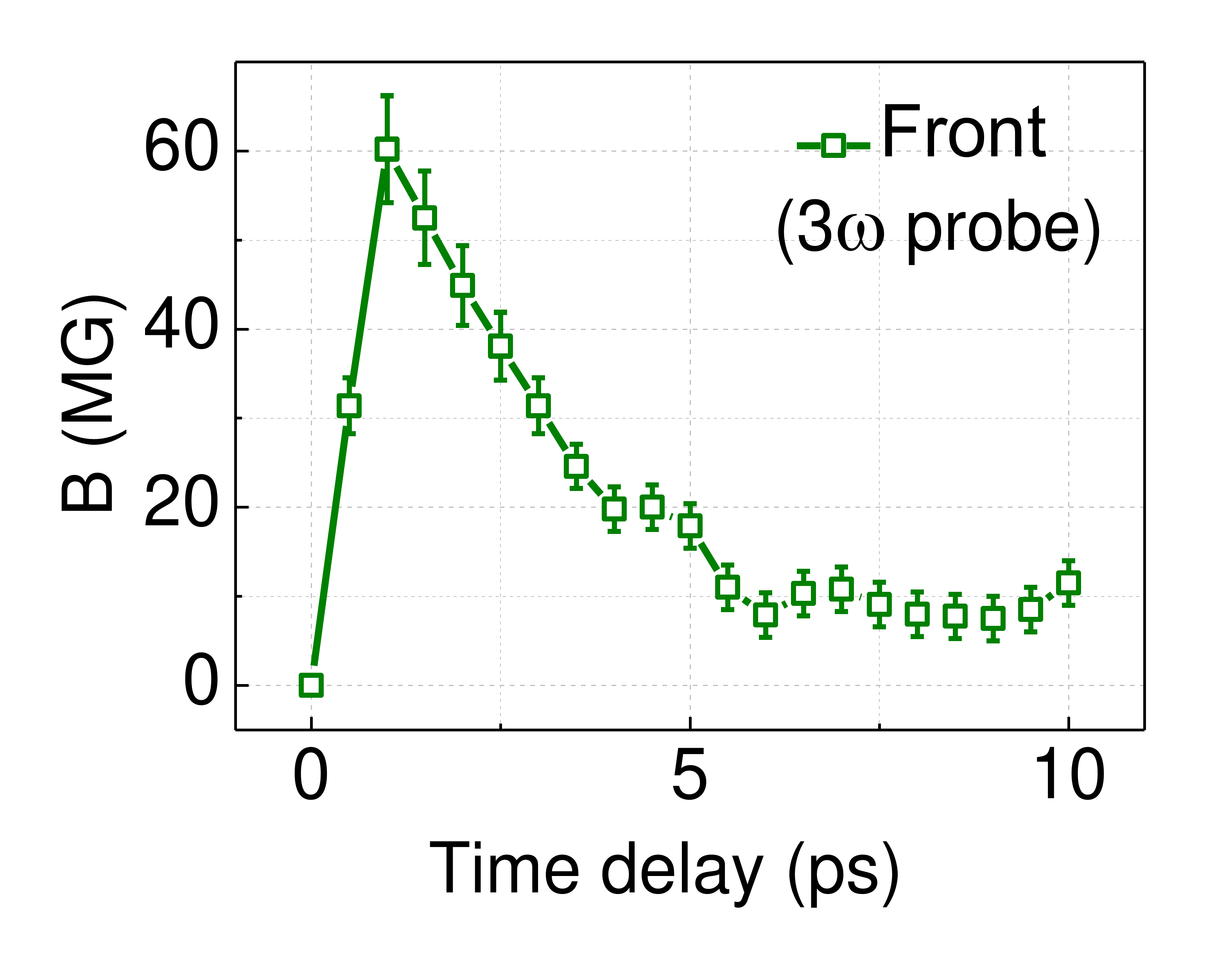}
\caption{\footnotesize{Temporal evolution of the spatially-averaged magnetic field at the front of an aluminum-coated thick BK7 glass target at an incident intensity of $3\times10^{18}$ W/cm$^2$ using a third harmonic probe.}} 
\end{figure}

Figure 11 clearly shows an initial sharp rise of the magnetic field at the target front, reaching a peak at $\sim 2$ ps, followed by a rapid decay until about 5 ps at a rate of ($1.4\pm0.3$) MG/ps and a subsequent slow decay at a rate of ($0.28\pm0.05$) MG/ps until 15 ps. At the target rear, the magnetic field follows a somewhat similar trend, reaching a peak of (2-3) MG at about 4 ps and exhibiting a more gentle growth rate of ($0.5\pm0.2$) MG/ps and a decay rate of ($0.28\pm0.04$) MG/ps, similar to that at the target front. A phenomenological modeling of the rise and decay time-scales of the magnetic field \cite{SandhuPRE2006} can provide an estimate of the conductivity of the bulk plasma as well as the penetration depth of the fast electrons (for measurements at the target front).

Figure 12 shows the temporal evolution of the spatially-averaged (across the transverse profile) magnetic field at the front of an aluminum-coated ($\sim 200-300$ nm thick) BK7 glass target at an irradiance of $3\times10^{18}$ W/cm$^2$ using a third harmonic probe. The average magnetic field is significantly higher, compared to that in Fig. 11. This is because of the aluminum-coating at the front surface, which results in a significantly more effective hot electron generation due to a lower ionization threshold of aluminum compared to glass, leading to a longer scale-length of interaction and a more efficient laser-energy coupling. In addition, a third harmonic probe penetrates deeper into the overdense plasma until a density of 9$n_c$, integrating the ellipticity it experiences due to the magnetic field along its entire path. The magnetic field appears as an ultrashort pulse, lasting for about 5 ps, as in Fig. 11, and decays rapidly at an initial rate of ($13\pm2$) MG/ps, followed by a slower decay rate of ($2.0\pm0.5$) MG/ps to reach asymptotic values of a few MG.  

In addition to shot-to-shot laser intensity fluctuations, the predominant sources of error in the measurement of ellipticity may be traced to various sources of extraneous noise as well as the depolarization induced by the optics in the probe imaging line, as discussed above. As a result, the base-level ellipticity may be non-negligible and hence ellipticity changes due to the magnetic field may be less prominent. Another significant source of error while mapping the magnetic field from the ellipticity is the propagated error from the estimation of the scale-length of the interaction, which is in turn a function of the volumetric heating of the plasma. Although significant at initial time-scales (about a picosecond) for the rapidly expanding plasma, at later time-scales, the magnetic field magnitude is not too sensitive to the scale-length and even a 100 eV change in the bulk plasma temperature typically results in only a few-megagauss change in the magnetic field.

\section{Summary} 
In summary, this paper recapitulates the basic working principle of Cotton-Mouton polarimetry of an externally launched probe for investigating ultrashort-pulse high-intensity laser-produced plasmas, with an emphasis on the advantages of this technique in measuring megagauss magnetic fields. We provide glimpses of the efficacy of this measurement technique by citing some of our previous experiments, aimed at investigating filamentary mechanisms, fast electron transport, and in general the rich physics involved in intense laser-plasma interactions. Measurements of the sub-picosecond temporal evolution and/or micron-scale spatial evolution of the megagauss magnetic fields have been the key diagnostic in all these experiments. In addition, we present recently obtained data using various innovations and improvements made in the measurement technique in order to expand the scope and application of these measurements -- an example being the simultaneous probing of the plasma up to near-solid densities using various probe harmonics, which provides a density-map of the magnetic fields. With a view to providing a consolidated modus-operandi, we have enumerated the various aspects of the measurement technique, based on our experience, which render the protocol easily implementable. In light of the pivotal role played by these megagauss magnetic fields in intense laser-plasma interactions in various disciplines, we hope our illustration of this measurement technique provides a platform for unraveling the various aspects of intense laser-plasma interactions which are not yet fully understood.

\section{Acknowledgements}
G.R.K. acknowledges financial support from a J. C. Bose grant (DST, Govt. of India). G.C. and P.K.S. acknowledge support from the Strong Field Science program (11P-1401). The authors thank M. Dalui, S. Tata, R. Gopal and J. Jha of TIFR for the SEQUOIA measurements. The authors gratefully acknowledge A. S. Sandhu, S. Kahaly, S. Mondal, V. Narayanan and Saima Ahmed for contributing to the development of this technique at TIFR.

\end{document}